# Exploring Climate with Obliquity in a Variable-eccentricity Earth-like World

M. J. Way[1,2,3] , Nikolaos Georgakarakos[4,5] , and Thomas L. Clune[6]
[1] NASA Goddard Institute for Space Studies, 2880 Broadway, New York, NY 10025, USA; Michael.Way@nasa.gov
[2] GSFC Sellers Exoplanet Environments Collaboration, NASA Goddard Space Flight Center, MD, USA
[3] Theoretical Astrophysics, Department of Physics and Astronomy, Uppsala University, Uppsala, SE-75120, Sweden
[4] Division of Science, New York University Abu Dhabi, P.O. Box 129188, Abu Dhabi, UAE
[5] Center for Astrophysics and Space Science (CASS), New York University Abu Dhabi, P.O. Box 129188, Abu Dhabi, UAE
[6] Global Modeling and Assimilation Office, NASA Goddard Space Flight Center, Greenbelt, MD 20771, USA
Received 2023 August 1; revised 2023 October 2; accepted 2023 October 13; published 2023 November 7

## Abstract

Exploring planetary systems similar to our solar system can provide a means to explore a large range of possibly temperate climates on Earth-like worlds. Rather than run hundreds of simulations with different eccentricities at fixed obliquities, our variable-eccentricity approach provides a means to cover an incredibly large parameter space. Herein Jupiter's orbital radius is moved substantially inward in two different scenarios, causing a forcing on Earth's eccentricity. In one case, the eccentricity of Earth varies from 0 to 0.27 over ∼7000 yr for three different fixed obliquities (0°, 23°, and 45°). In another case, the eccentricity varies from 0 to 0.53 over ∼9400 yr in a single case with zero obliquity. In all cases, we find that the climate remains stable, but regional habitability changes through time in unique ways. At the same time, the moist greenhouse state is approached but only when at the highest eccentricities.

*Unified Astronomy Thesaurus concepts:* Exoplanet astronomy (486); Exoplanet atmospheres (487); Exoplanet dynamics (490)

## 1. Introduction

Since the discovery of the first exoplanet nearly three decades ago, another 5000+ exoplanets have been found. These planets have been found in a variety of systems (single-star, multi-stellar, multi-planet) and orbital architectures. The search for a habitable planet in one of these systems is a key goal within the scientific community. This question has become more relevant since the detection of planets with masses and sizes similar to those of Earth is now common. In addition, some of these planets (e.g., Gillon et al. 2016; Gilbert et al. 2020, 2023; Kossakowski et al. 2023) are found in the habitable zone of their host stars.

Planetary habitability is a multifaceted problem that requires an understanding of several properties and characteristics of the host star, the planet, the planetary system it belongs to, and its location within its host galaxy (e.g., see Meadows & Barnes 2018 for a review). In our solar system, the orbits of the planets are almost circular and coplanar. Mercury has the highest eccentricity of around 0.2 and an orbital inclination of about 7°. Some exoplanets have been found to move on highly eccentric orbits with HD2082b being the most eccentric exoplanet discovered thus far at $e = 0.97$ (Stassun et al. 2017). Gliese 514b is a potentially habitable planet with a 3$\sigma$ upper limit eccentricity of 0.9 (Damasso et al. 2022). A number of systems have also been discovered to have noncoplanar orbits (e.g., McArthur et al. 2010). The eccentricity of the planetary orbit is an important factor as it affects the instellation the planet receives from its host star. In planetary systems with more than two bodies, the eccentricity may oscillate on various timescales and with significant amplitudes, while the semimajor axis shows little to almost no variation (e.g., see Georgakarakos 2003; Georgakarakos & Eggl 2015). This behavior of the planetary orbit may affect the potential of a planet to retain liquid water on its surface (e.g., Way & Georgakarakos 2017; Georgakarakos et al. 2018).

The planetary obliquity, i.e., the angle between the planetary spin axis and the normal of its orbital plane, is another important factor that may affect the climate of the planet. It has been the subject of many investigations either at the purely dynamical level, i.e., how it evolves under the gravitational perturbations of other bodies in the system (e.g., Atobe et al. 2004; Saillenfest et al. 2019) or with respect to habitability and climate evolution. An example of the latter type of work is Williams & Kasting (1997), who simulated the climate of Earth at different obliquities using a 1D energy balance model. Williams & Kasting (1997) and Spiegel et al. (2009) used the same model to investigate the climate of a terrestrial planet at various obliquities. By using a 3D atmospheric general circulation model (GCM), Wang et al. (2016) studied the effect of obliquity on the habitability of planets around around M dwarf stars but with a lower-complexity thermodynamic 50 m slab ocean with zero horizontal heat transport. Additional studies have examined the relationship between obliquity and planetary climate (e.g., Ferreira et al. 2014; Kilic et al. 2017; Rose et al. 2017; Colose et al. 2019). Armstrong et al. (2014) explored habitability by using an energy balance model and simulating the effects of orbital and obliquity evolution of terrestrial planets in hypothetical systems. Williams & Pollard (2002) examined climate on Earth-like worlds with an energy balance model (EBM) and a GCM. In later work, Williams & Pollard (2003) again use a GCM, but in both of their works they used a thermodynamic 50 m slab ocean (as in Wang et al. 2016). Finally, Linsenmeier et al. (2015) used a GCM for different values of planetary eccentricity and obliquity to explore the effects of seasonal variability on the climates of Earth-like planets, but as in the previous GCM studies

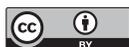

mentioned above, they use a thermodynamic 50 m slab ocean with zero horizontal heat transport.

Previously, we explored the climate evolution of an Earth-like planet around a Sun-like star under the gravitational perturbations of a giant planet (Way & Georgakarakos 2017) using the ROCKE-3D GCM (Way et al. 2017). The idea behind that study was to quantify the effects of the planet's variable eccentricity on its climate while maintaining the modern value of Earth's obliquity. In this work, we extend our previous study by using different obliquity values and an additional simulation with a larger variable eccentricity range. A key difference between our previous/present work and that of other similar GCM studies is that we utilize a fully coupled dynamic ocean (e.g., Way et al. 2017, Section 2.2), not a thermodynamic slab ocean. We discuss the methods in Section 2, discussion in Section 3, and conclusions in Section 4.

## 2. Methods

In Way & Georgakarakos (2017) we explored two different planetary configurations. For the first configuration, a terrestrial planet was placed on an orbit with a semimajor axis equal to 1 au from a Sun-like star, while a Jupiter mass body orbited the star at $a_J = 1.8$ au on a slightly eccentric orbit with $e_J = 0.05$. In the second setup, the Jupiter-like body was given a semimajor axis of $a_J = 2.15$ au with an eccentricity of $e_J = 0.27$. In both scenarios, the obliquity of the terrestrial planet was fixed to that of modern Earth. All bodies resided in the same plane of motion. The idea behind varying a single parameter in this manner was that we desired to have systems that would produce variations in the eccentricity and hence in the distance between the star and the terrestrial planet, which would result in changes in the incoming radiation. The orbital evolution of the Earth-like world was followed using the analytical model of Georgakarakos et al. (2016).

In this work, we have extended the results of Way & Georgakarakos (2017). We keep the second configuration from that work, but we modify the obliquity of the terrestrial planet. We run two additional cases of 0° and 45° polar obliquity. On top of that, we run a new configuration of a Jupiter mass body at $a_J = 2.4$ au with an eccentricity of $e_J = 0.4$. That is done in order to study a configuration where the terrestrial planet will acquire an eccentricity as large as $e_{TP} = 0.53$. The obliquity for this dynamical scenario was set to $o_{TP} = 0°$. In this case, the orbital evolution was obtained through the numerical integration of the full equations of motion of the system. This was done because the analytical model may not provide entirely accurate results due to the higher values of the terrestrial planetary eccentricity $e_{TP}$. The integration time for both systems was set to one secular period, i.e., 7000 yr for the systems where $a_J = 2.15$ au and 9418 yr for the systems with $a_J = 2.4$ au. The orbital configurations selected for this study are dynamically stable over long timescales. The empirical stability formula of Petrovich (2015) gives a critical semimajor axis of $a_{TP} = 1.17$ au for the terrestrial planet in the system where the Jupiter mass body has $a_J = 2.15$ au, while for the other system, the critical semimajor axis is $a_{TP} = 1.12$ au. Numerical simulations of the full equations of motion of both systems for at least 100 secular periods confirmed their long-term dynamical stability. Such timescales should be long enough for our three-body coplanar systems to demonstrate any potential orbital changes due to gravitational interactions that may take longer to show (e.g., secular resonances). The stability simulations revealed no orbital changes that would cause concern for the long-term stability of the systems under investigation.

In this study, we again utilize a fully coupled ocean and atmosphere GCM called ROCKE-3D (Way et al. 2017). We model a world very similar to modern Earth as detailed in Way et al. (2018). The major differences from modern Earth include:

1. Ocean depth is fixed at 1360 m except at continental margins where it is set to 591 m.
2. Large seas are removed, e.g., the Baltic, Hudson Bay, the Mediterranean, and the Black and Caspian Seas.
3. The strait that separates Greenland from northern Canada is slightly expanded.

These changes were mainly implemented because

1. A shallower-than-modern-Earth ocean will come into equilibrium faster, yet at 1390 m (the depth of most ocean/sea grid cells) it still has a great deal of thermal inertia.
2. Removing shallow ocean or sea grid cells avoids cases where they may freeze to the bottom, which crashes the model because it cannot dynamically change surface types (in this case from ocean to ground ice).

In Cases 1–3, we fixed the longitude of periapsis (LoP) to modern-day Earth's value of 282°9 even though our dynamical calculations provided this number evolving with the eccentricity. This was done to make the interpretation of results simpler in that the seasons driven by obliquity in modern Earth remain roughly in sync with the perihelion and aphelion passage. On the other hand, in Case 4, we utilized the variable LoP. The difference between fixing LoP and the variable LoP is clearly seen in the zero-obliquity cases (1 and 4) shown in Figure 1. Here, Case 1 seasonal insolation remains like that of modern Earth (Column 2), while Case 4 has shifted. For Case 2, the insolation patterns are obviously that of modern Earth, but had we run a version of Case 4 with $o_{TP} = 23°5$, some of the symmetries we see in the seasonal patterns would clearly be shifted. The insolation patterns in Figure 1 are favorably comparable with some of those in Dobrovolskis (2013) and Jernigan et al. (2023).

For Cases 1–3 (see Table 1) the aphelion distance results in installations as low as 844 W m$^{-2}$ or 62% of modern Earth's insolation and 571 W m$^{-2}$ or 42% of modern Earth for Case 4. Even if these are over very short periods of time (i.e., a single Earth year) it is possible for shallow grid cells to freeze to the bottom, which may crash the GCM. The land sea mask and bathymetry changes mentioned above prevent such crashes from occurring. Modern-Earth atmospheric compositions are used: 77% $N_2$, 21% $O_2$, and preindustrial (1850) amounts for the major greenhouse gas components $CO_2 = 285$ ppmv, $N_2O = 0.27$ ppmv, and $CH_4 = 0.79$ ppmv. In addition, modern-Earth $O_3$ values are retained. The native GISS radiation scheme is used in Cases 1, 2, and 3. This radiative transfer scheme is relatively fast, and since the model stays within the radiation's permitted range of temperatures, it is appropriate to use it for these cases. For Case 4, two major differences are that $O_3$ and aerosols are omitted (lack of aerosols will tend to make the model warmer, and there will be no $O_3$-associated tropopause cold trap), and we utilize the SOCRATES[7] (Edwards 1996; Edwards & Slingo 1996) radiative transfer

---

[7] Suite of Community Radiative Transfer codes based on Edwards and Slingo.

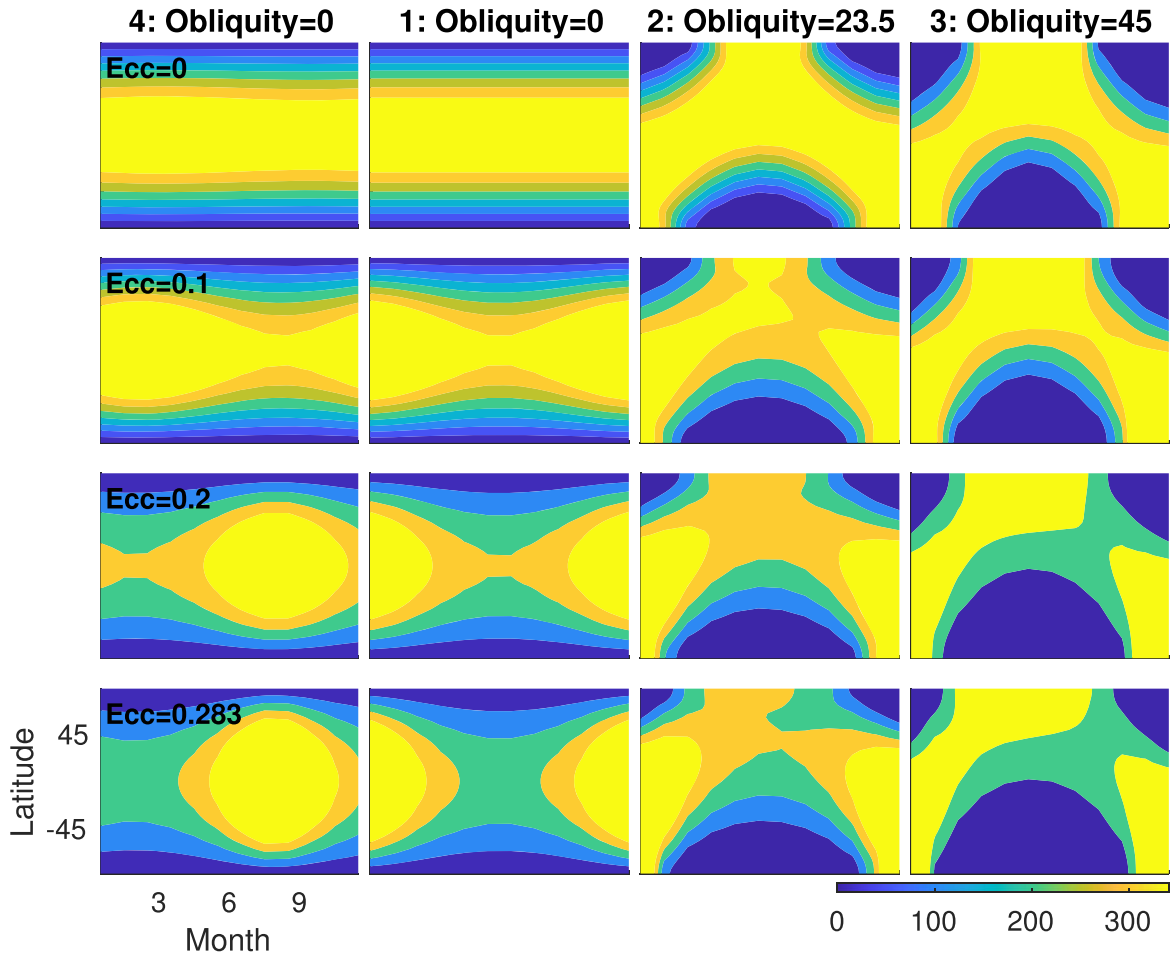

**Figure 1.** Comparable monthly insolation patterns for Cases 1–4 at given snapshots in their eccentricity evolution. We do not include snapshots of the higher eccentricity values attained in Case 4. Insolation bar in the lower right goes from zero to 341 W m$^{-2}$. Note that we plot Case 4 in the first column and Cases 1–3 in columns 2–4 to make comparisons between Cases 4 and 1 easier.

**Table 1**
Simulations

| | Jupiter | | Earth | | | | | Run Time | | Ins[a] |
|---|---|---|---|---|---|---|---|---|---|---|
| Case | Eccentricity | Semimajor Axis | Eccentricity Variable | Obliquity (deg) | Orbital/Rotation Period (day/hr) | Semimajor Axis | Longitude[b] Periapsis | Model (yr) | Wall Clock (days) | (min/max) |
| 1 | 0.27 | 2.15 | 0–0.27 | 0 | 365/24 | 1.00 | Fixed/Modern | 7000 | 108 | 0.62/1.84 |
| 2 | ,, | ,, | ,, | 23 | ,, | ,, | ,, | ,, | ,, | ,, |
| 3 | ,, | ,, | ,, | 45 | ,, | ,, | ,, | ,, | ,, | ,, |
| 4 | 0.4 | 2.4 | 0–0.53 | 0 | ,, | ,, | Variable | 9418 | 660 | 0.42/4.20 |

**Note.** Summary of the simulation setups. Ins[a]: instellation as fraction of modern Earth's 1361 W m$^{-2}$ insolation. LP[b]: LoP is either fixed to modern Earth's value of 282°9 (Cases 1–3) or allowed to evolve through time (Case 4).

scheme, which allows for a larger range of temperatures than the native GISS scheme. Higher temperatures were expected with the higher insolations in Case 4, which can be seen in Figure 3 where global mean surface temperatures reached 69°C. Both GISS and SOCRATES radiative transfer schemes are described in detail in Way et al. (2017). The speed difference between the radiation scheme is born out in the run time "wall clock" column of Table 1 where Cases 1–3 were each completed in ∼108 days (7000/108 ∼ 65 model yr day$^{-1}$), whereas Case 4 took 660 days (9418/660

∼14 model yr day$^{-1}$). Hence the SOCRATES scheme runs nearly 4.6 times slower than the GISS scheme (64/14 ∼ 4.6). The slowness of SOCRATES and the length of time to complete Case 4 mean that we only ran a single case with zero obliquity. The time and resources required to complete another two obliquity cases to quantitatively compare with Cases 1–3 were inhibiting. In this sense, Case 4 is more a proof of concept, and our limited discussion below reflects this fact.

For Cases 1–3, the baseline model was modified by introducing a new function that computes the custom eccentricity as a function of time. This function was then invoked daily to update the orbital parameters. For Case 4, our changes were somewhat more involved. Here, we implemented a new subclass, ForcedOrbit, which extends and customizes a ROCKE-3D function called PlanetaryOrbit. In this extension, the usual procedure for updating the orbital angles (longitude, decl., and hour angle) is modified to periodically check a file to update the primary orbital parameters. The file contains updates for distance from primary, mean anomaly, eccentricity, and longitude of pericenter. The longitude of pericenter is used to adjust the longitude at periapsis from its initial value. The customized model source code for both of these modifications can be obtained.[8]

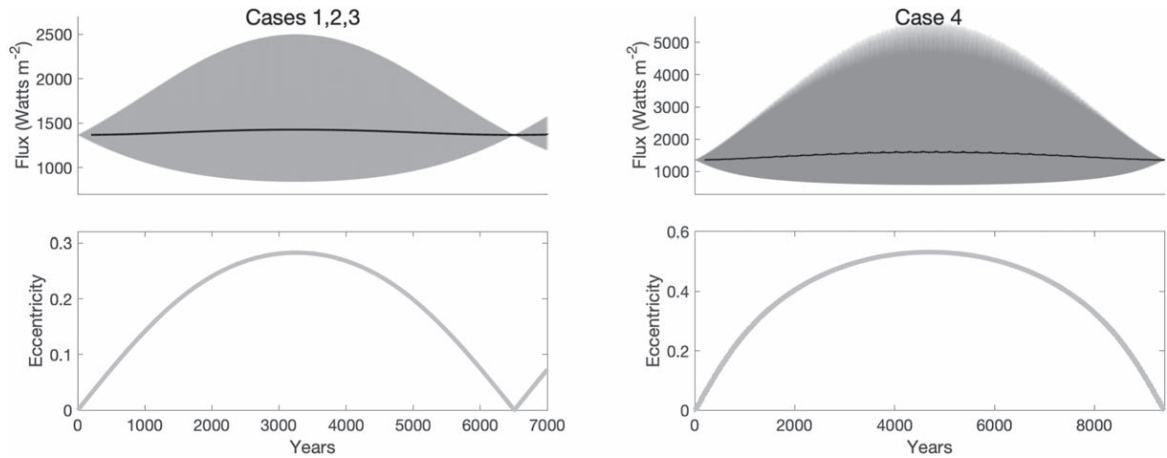

**Figure 2.** Instellation (Flux W m$^{-2}$) and eccentricity through time for Cases 1, 2, 3 (left) and Case 4 (right). Black line in upper panels is a 120 month moving mean. Note that $y$-axes have different scales. The $x$-axis scales are different in the left vs. right set of plots. See Table 1 for details.

## 3. Discussion

Below we discuss a few critical GCM diagnostics to better understand how the models evolve through time. Figures 3–5 show relevant diagnostic GCM output for our analysis. The figures in Section 3 are 120 months running means of each diagnostic. See Appendix (Figures A1–A3) where the full data are plotted along with the 120 month mean.

### 3.1. Radiative Equilibrium

In Figure 3 we first look at the net radiative balance (subplot (D)) at the top of the atmosphere. Generally the ROCKE-3D model is considered in balance when this number averaged over 10 yr is between $\pm 0.2$ W m$^{-2}$ (as described in Way et al. 2017). The thin dark black line is a 120 month (10 yr) running average. Although perhaps not apparent in this plot given the limits required to see the full data set, the model is seldom within the desired $\pm 0.2$ range. However, this is expected since we are in essence constantly changing the forcing at the top of the atmosphere through time via our changing eccentricity (see Figure 2). In general, the model is $< 1$ W m$^{-2}$ out of balance at any given time for Case 1. for Cases 2 and 3, this increases to $<3$ W m$^{-2}$. These numbers are quite acceptable given the eccentricity forcing through time and the variability caused by ever-large obliquities from Case 1 to Case 3. In Case 4 with the higher eccentricities modeled, the model imbalance is as large as 9 W m$^{-2}$, but in general averages over 1000 yr are around 3–4 W m$^{-2}$ for this case. Again, this is not unreasonable given the constantly changing solar forcing on the atmosphere. The thermal capacity of the ocean is large in comparison with the atmosphere, and hence, just as on Earth where the hottest and coldest months of the year are 1–2 months after summer and winter solstice, the same rule applies here. Hence, we do not expect the models to maintain radiative equilibrium due to the constantly changing solar forcing at the top of the atmosphere.

### 3.2. Albedo

Ground or surface albedos are particularly important to estimate given the reliance of the community on them for equilibrium temperature calculations (e.g., Genio et al. 2019) and the manner in which they are calculated in 1D models of the habitable zone (e.g., Kasting et al. 1993; Kopparapu et al. 2013).

The planetary albedo can be affected by surface changes through time via increases/decreases in ground or sea ice, soil water saturation (wetter soils tend to be darker than the same dry soils), and of course, clouds. In Figure 3(B), it is clear that the lower obliquities (Cases 1 and 4) correlate with higher planetary albedos. This is also reflected in the maximum values in Table 2.

In Figure 4, the ocean ice fraction percentage is higher in all years for Case 1 versus the higher-obliquity Case 2 and 3. This is due to the fact that the poles are receiving less insolation in the zero-obliquity case. This is also seen in Figure A4 for Cases 1–3 where the maximum extent of oceanic ice sheets are correlated with obliquity. However, this general pattern does not hold in Case 4 with zero obliquity. This is probably because the higher eccentricities generate such high insolation at aphelion that the ice fraction remains surprisingly low as seen in Figures 4 and A4. However, in Figure 4, one may notice that toward the end of the Case 4 run (year ~9400) the ocean ice fraction appears to climb to values approaching 15%, rather than the 11%–12% toward the beginning of the run (year ~400). Hence, if we were able to run the model longer, we might see behavior more similar to that of Case 1 given the possible strength of the ice albedo feedback.

Ground albedo (Figure 5(C)) is a combination of snow+ice coverage (subplot (a)), and perhaps a small amount from soil/ground water saturation (subplot (b)). Moving from Case 1 to 3, one sees decreasing Snow+Ice fractions. This is again due to the lower obliquity cases having more persistent ice in polar

---

**Table 2**
Maximum and Minimum Global Diagnostics

| Case | Temperature (Celsius) | Albedo (%) | Grnd Alb (%) | Precipitation (mm day$^{-1}$) | 100mb S.H. (v v$^{-1}$) | Surface S.H. (kg kg$^{-1}$) | Ocean Ice (%) | Snow Depth (m) | Clouds (%) |
|---|---|---|---|---|---|---|---|---|---|
| 1 | 5.3/24.2 | 30.2/36.4 | 9.0/19.9 | 2.3/3.8 | $3.0 \times 10^{-6}/4.4 \times 10^{-5}$ | 0.007/0.019 | 0/16.9 | 0.007/0.300 | 56/67 |
| 2 | 8.4/25.8 | 28.6/34.0 | 8.8/18.4 | 2.4/3.9 | $2.9 \times 10^{-6}/4.0 \times 10^{-5}$ | 0.007/0.018 | 0/8.9 | 0.006/0.169 | 54/66 |
| 3 | 11.8/28.0 | 27.0/34.1 | 8.6/14.4 | 0.9/4.1 | $4.7 \times 10^{-6}/5.9 \times 10^{-5}$ | 0.009/0.020 | 0/4.2 | 0.005/0.050 | 54/67 |
| 4 | 8.3/69.1 | 21.3/46.3 | 7.9/16.4 | 0.7/7.1 | $2.5 \times 10^{-10}/7.2 \times 10^{-3}$ | 0.008/0.098 | 0/15.2 | 0.004/0.324 | 33/86 |

**Note.** Minimum and maximum global diagnostic values over length of each simulation. Grnd Alb = Ground Albedo, S.H. = Specific Humidity.

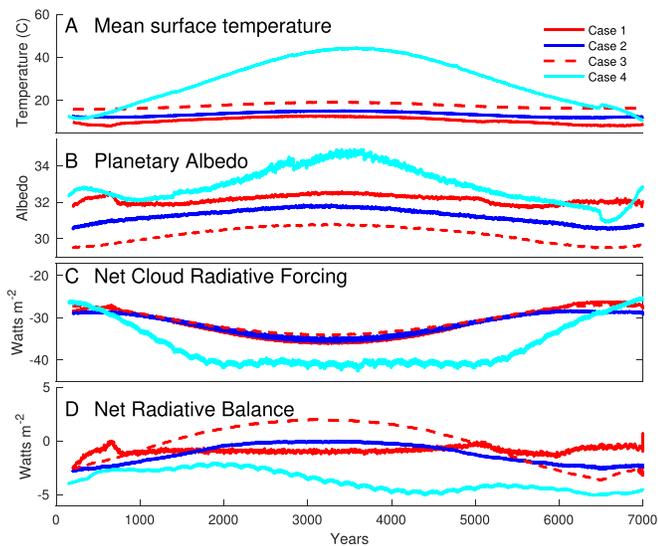

**Figure 3.** Global diagnostics: surface temperature, planetary albedo, net cloud radiative forcing, and net radiative balance. Lines are 120 month moving means. Note that to make all four cases easily comparable, we have scaled the number of years (x-axis) in Case 4 to the same as Cases 1–3. Nothing changes in Case 4 in the y-axis.

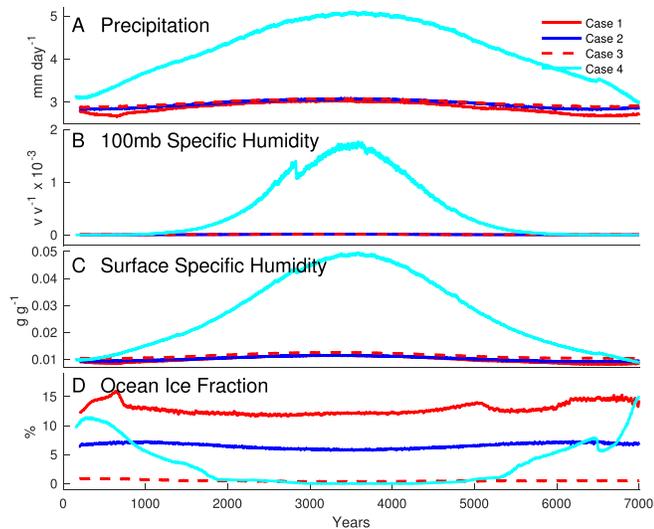

**Figure 4.** Global diagnostics: precipitation, top of the model (100 mb) SH in volume mixing ratios (H$_2$O/Air), surface SH in mass mixing ratios (H$_2$O/Air), and ocean ice fraction (in percent). Lines are 120 months moving means. Note that the Kasting et al. (1993) moist greenhouse limit of $3 \times 10^{-3}$ v v$^{-1}$ is not reached for any run although see Figure A2. Note that to make all four cases easily comparable, we have scaled the number of years (x-axis) in Case 4 to the same as Cases 1–3. Nothing changes in Case 4 in the y-axis.

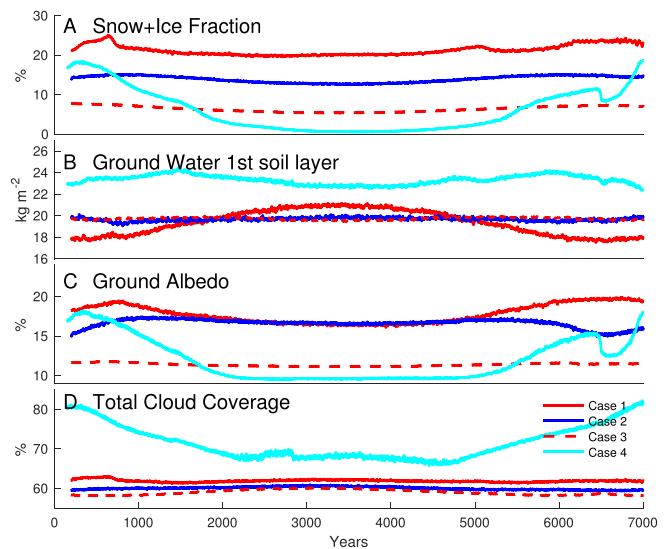

**Figure 5.** Global diagnostics: snow+ice fraction, ground water amount in the first soil layer, ground albedo, and total cloud coverage. Lines are 120 month moving means. Note that to make all four cases easily comparable, we have scaled the number of years (x-axis) in Case 4 to the same as Cases 1–3. Nothing changes in Case 4 in the y-axis.

regions. This is partially reflected in the ground albedo (subplot (c)). One might expect the increases in the water amounts in the top layer of the soil (subplot (b)) to lower the ground albedo as wet soil has a lower albedo than dry soil. That correlation is seen in Case 1 when looking at the trend of Figures 5(B) and (C). The higher amounts of ground water in the first layer may be influencing the ground albedo, driving it slightly lower during years of higher eccentricity, and it may correlate with the larger precipitation seen in Figure 4(A) (see Figure A2 for more detail). At the same time, wider ranges in annual precipitation is seen in all models during periods of lower eccentricity (Figure 4(A)) but decreases slightly from low to high obliquity. Cases 1 and 4 follow similar trends in ground albedo at higher eccentricities where they have lower ground albedos in these two zero-obliquity cases. This is due to relatively lower snow-ice fractions at higher eccentricities in Cases 1 and 4 (Figure 5(A)).

Cloud fractional coverage, or what we term total cloud coverage (Figure 5(D) and Figure A3) shows some correlation with planetary albedo (Figure 3 (B)) for Cases 1–3. Precipitation follows a similar trend regardless of obliquity (Figure 4 (A)) although Case 3 has larger annual variability at lower eccentricities than either Case 1 or 2 (Figure A2 in the Appendix). Precipitation in Case 4 follows a similar trend with higher precipitation during higher eccentricities. But note that Case 4 has more extremes in annual cloud coverage with values as large as 86% (at lower eccentricities) and as low as 33% (at the highest eccentricities) as shown in Figure A2 and Table 2.

Our study could provide useful constraints for EBM models like that of Haqq-Misra et al. (2016) and Ji et al. (2023) where the albedo may be set to a constant. Here it is clear from the planetary albedo (Figures 3 and A1) and ground albedo (Figures 5 and A3) that this may not be a good approximation (see Table 2 for the min./max. values).

### 3.3. Surface and Upper-atmosphere Specific Humidity

First we examine the specific humidity (SH) at 100mb; in other words, how wet is the stratosphere? As pointed out in the work of Kasting (1988), a value of SH greater than $10^{-3}$ (v v$^{-1}$), later increased to $3 \times 10^{-3}$ in Kasting et al. (1993 Section 5(i)), is considered undesirable for long-term habitability if hydrogen escapes at the diffusion limit as an entire Earth's ocean could be lost in less than 4 Gyr. The 100mb SH never approaches $3 \times 10^{-3}$ even at the highest eccentricities for Cases 1–3 in Figure 4. For Case 4, Figure 4 shows it also remains below the Kasting limit in the running mean. However, Figure A2 shows that during periods of high eccentricity, there are periods of time at aphelion where the limit is surpassed for Case 4. The latter is in line with recently published work by Liu et al. (2023).

Examining surface SH, the biggest concern is whether values approach 10%. At that level water, becomes a non-negligible component of the atmosphere and may begin to affect the mean molecular weight of the atmosphere that is set at the start of each run and is not allowed to change. This in turn would affect the weight of air parcels, and hence, convection will begin to diverge from reality. Only in Case 4 (Figure A2) does the model begin to approach 10% at the highest eccentricities during aphelion (in the running mean, it remains well below this limit as shown in Figure 4) but mostly remains below this critical value and hence presents none of the concerns raised herein.

### 3.4. Atmospheric and Oceanic Meridional Transport

Meridional transport plays a critical role in moving heat from lower to higher latitudes in planets like Earth (e.g., Masuda 1988; Trenberth & Caron 2001) and those with the modest obliquities modeled herein ($o_{TP} = 0°$, 23°.5, and 45°). In Figure 4 it is clear that lower obliquity worlds tend to have more ocean ice at higher latitudes (see Figure A4). Interestingly, the system works hard to move heat poleward for lower obliquities regardless of eccentricity. This can clearly be seen in Figure A5 where the lower obliquity runs for Cases 1 and 4 are transporting more energy toward the poles than the higher-obliquity Cases 2 and 3. The trend continues regardless of eccentricity where Case 2 ($o_{TP} = 23°.5$) also transports more heat poleward than Case 3 ($o_{TP} = 45°$).

In fact, it is the winds that drive the ocean currents and the meridional overturning circulation as pointed out in previous work (e.g., Ferreira et al. 2014). In Figures A6(A) and (B), when looking at an annual mean, one can see that indeed the mean wind speeds are much higher in Case 1 ($o_{TP} = 0°$) at 3.5 m s$^{-1}$ versus Case 3 ($o_{TP} = 45°$) at 1.7 m s$^{-1}$ (keeping eccentricity = 0 in both figures). To further illustrate the point, Figure A6(C) plots the wind speeds in Case 1 (for eccentricity = 0) in the month of January versus Case 1 in January at eccentricity = 0.283 (Figure A6(D)). January is chosen since its near perihelion when insolation is at its highest in the southern hemisphere where there is more ocean than land and the oceanic meridional circulation is strongest. The wind speeds are only slightly different with the mean and max of the former (Case 1 eccentricity = 0) being 3.7 and 12.1 m s$^{-1}$, while the latter (Case 1 eccentricity= 0.283) has 4.1 and 11.8 m s$^{-1}$. For comparison purposes, Figures A6(D), (E) at aphelion (July) Case 1 at eccentricity= 0 has a mean of 3.6 and max of 10.7, while Case 1 at ecc = 0.283 has a mean of 3.3 and max of 11. Figure A6 (C), (E) (Case 1) should in theory have the same values since insolation at "perihelion" and "aphelion" are the same, but there are fluctuations in the system over time, and the global mean values are within 0.1 m s$^{-1}$ of each other as expected.

## 4. Conclusion

We have modeled four variable-eccentricity Earth-like worlds with obliquities ranging between 0° and 45°. These unique simulations provide an opportunity to examine a particularly wide parameter space in detail without having to generate a large parameter ensemble. We have examined a variety of model output diagnostics to compare and contrast these worlds and better understand the roles of eccentricity and obliquity on their climates. In general, we find a fully coupled dynamic ocean appears to provide a buffer keeping the climate relatively temperate as the simulations move to higher eccentricities with quite extreme insolations at aphelion and perihelion (see Table 2). This work may also have implications for exoplanetary systems. For example, it is traditionally assumed that most close-in terrestrial planets around M dwarfs are tidally locked with moderate eccentricities. However, this may not always be the case (e.g., Makarov et al. 2018), and work like this is an important step in characterizing such systems.

### Acknowledgments

This work was supported by NASA's Nexus for Exoplanet System Science (NExSS) and the NASA Interdisciplinary Consortia for Astrobiology Research (ICAR). Resources supporting this work were provided by the NASA High-End Computing (HEC) Program through the NASA Center for Climate Simulation (NCCS) at Goddard Space Flight Center. M.J.W. acknowledges support from the GSFC Sellers Exoplanet Environments Collaboration (SEEC), which is funded by the NASA Planetary Science Division's Internal Scientist Funding Model, and ROCKE-3D which is funded by the NASA Planetary and Earth Science Divisions Internal Scientist Funding Model. Data used to generate the figures herein can be downloaded from doi:10.5281/zenodo.8398270, while the custom GCM modifications used in ROCKE-3D version Planet_1.0 can be downloaded from the same doi alongside any necessary input files.

## Appendix
### Additional Figures

Figure A1 shows the polar ice extent for Cases 1-4 through time. Figure A2 demonstrates differences in total meridional transport (atmosphere + ocean) instances in Cases 1-3 eccentricity evolution, 0.0, 0.1, 0.2, and 0.28, and 0.0 to 0.53 for Case 4. Figure A3 shows global GCM diagnostics through time (a) mean global surface temperature; (b) planetary albedo; (c) net cloud radiative forcing; and (d) net radiative balance. Figure A4 shows diagnostics through time (a) global mean precipitation; (b) specific humidity at 100mb in the upper atmosphere; and (c) specific humidity at the surface and the global mean ocean+lake ice fraction. Figure A6 has additional global mean diagnostics (a) snow+ice fraction; (b) the amount of ground water in the first soil layer (the model has 6 soil layers); (c) the global mean ground albedo; and (d) total cloud coverage. Figure A6 contains the surface wind speed for annual means (a) Case 1 (zero obliquity) when at zero eccentricity, as a comparison with (b) Case 3 (45 degrees obliquity) when also at zero eccentricity. Next are shown January/perihelion wind speeds for (c). Case 1 at zero eccentricity versus (d) Case 1 at its highest eccentricity of 0.283. Finally wind speeds are shown at July/aphelion for Case 1 at (e) zero eccentricity and F. 0.283 eccentricity. The larger context for each set of figures are discussed in detail in the main text.

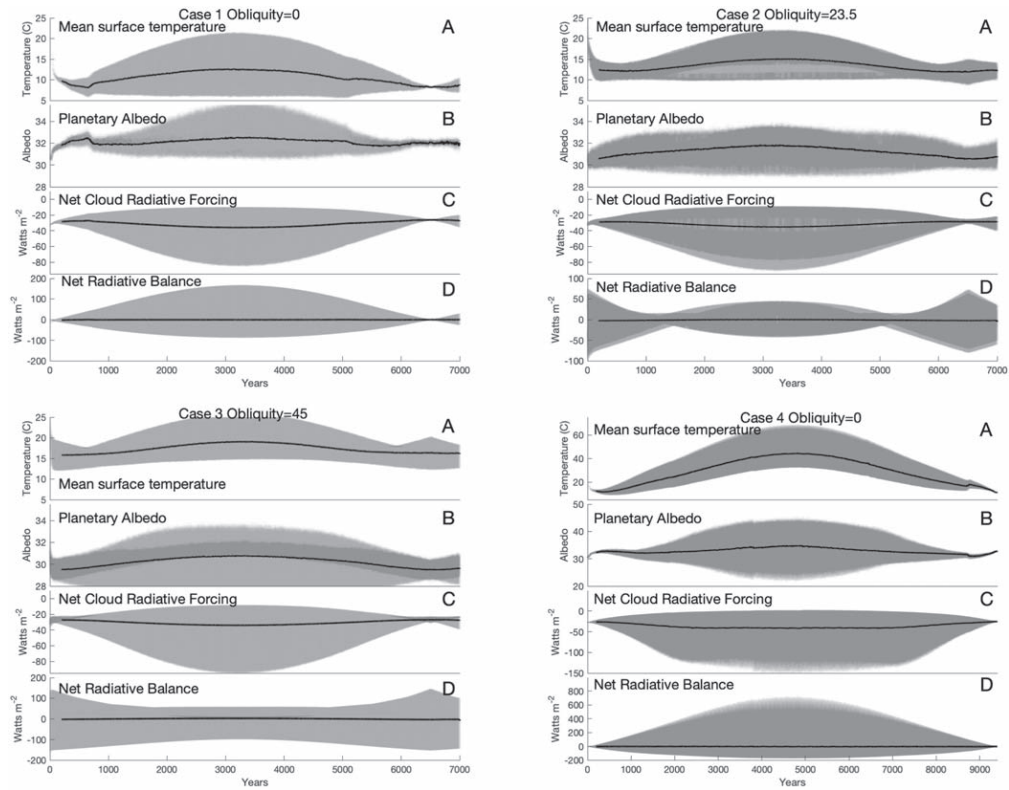

**Figure A1.** Global diagnostics: surface temperature, planetary albedo, net cloud radiative forcing, and net radiative balance. Black solid lines are 120 month moving means. Notice that Case 4 often has different x- and y-axis scaling from that of Cases 1–3.

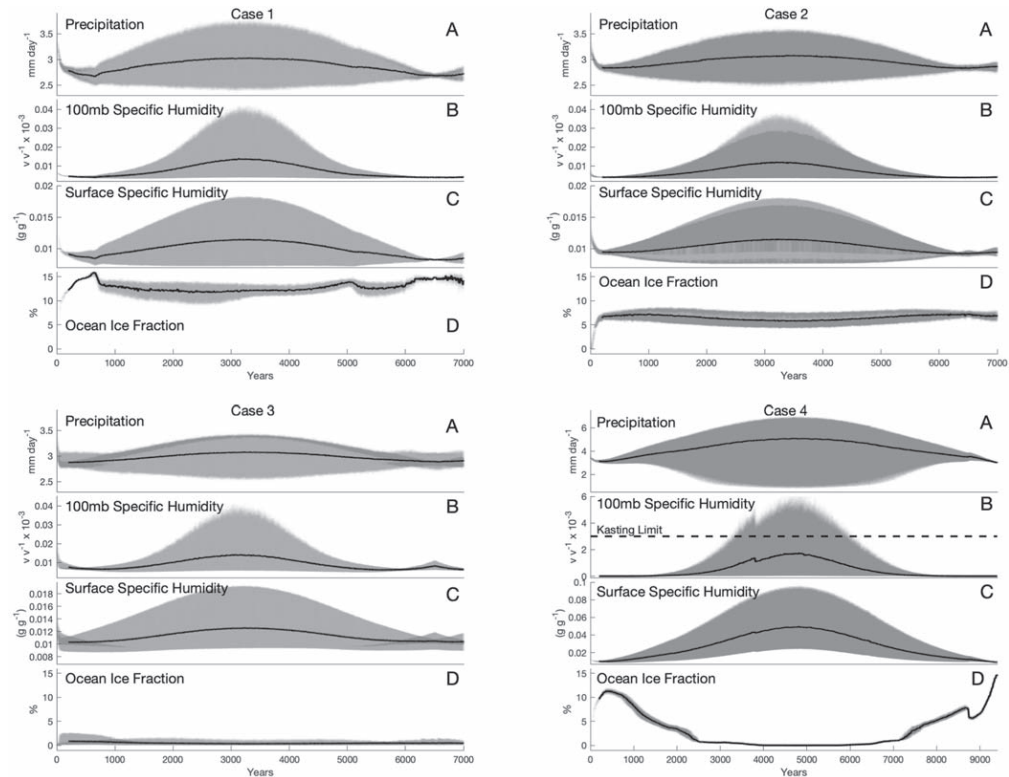

**Figure A2.** Global diagnostics: precipitation, top of the model (100mb) SH in volume mixing ratios (H$_2$O/Air), surface SH in mass mixing ratios (H$_2$O/Air), and ocean ice fraction (in percent). Black solid lines are 120 month moving means. The Kasting et al. (1993) moist greenhouse limit of $3 \times 10^{-3}$ v v$^{-1}$ is denoted by a black dashed line. Notice that Case 4 often has different x- and y-axis scaling from that of Cases 1–3.

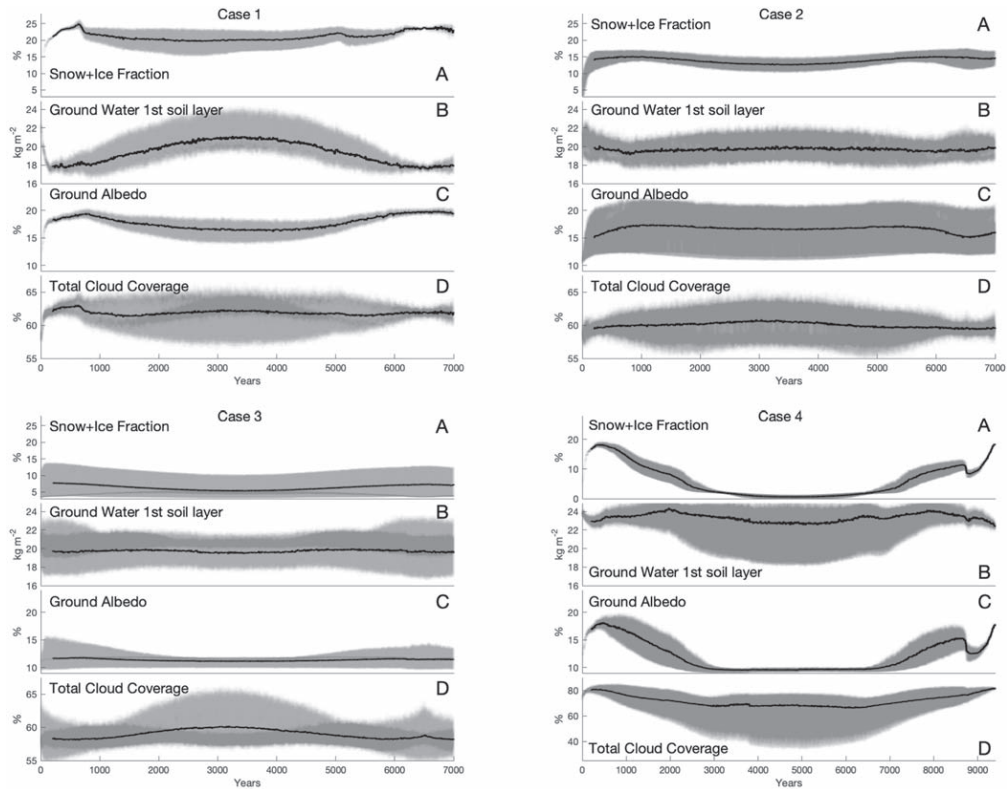

**Figure A3.** Global diagnostics: snow+ice fraction, ground water amount in the first soil layer, ground albedo, and total cloud coverage. Black solid lines are 120 month moving means. Notice that Case 4 often has different *x*- and *y*-axis scaling from that of Cases 1–3.

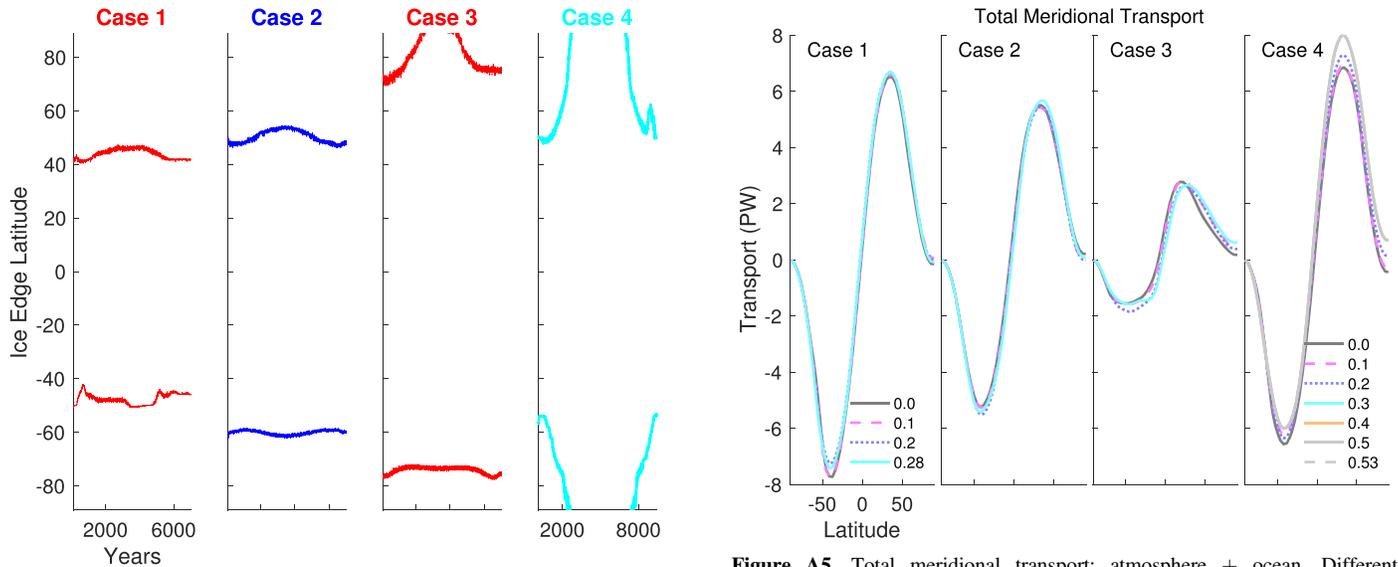

**Figure A4.** Maximum northern and southern latitudes of oceanic ice sheets. Note that part of the irregular patterns are related to the continents and their influence.

**Figure A5.** Total meridional transport: atmosphere + ocean. Different obliquities are plotted (Case 1–4) with snapshots (yearly means) at different eccentricities. Here we see the critical role played by obliquity for meridional transport, whereas eccentricity plays a less prominent role.

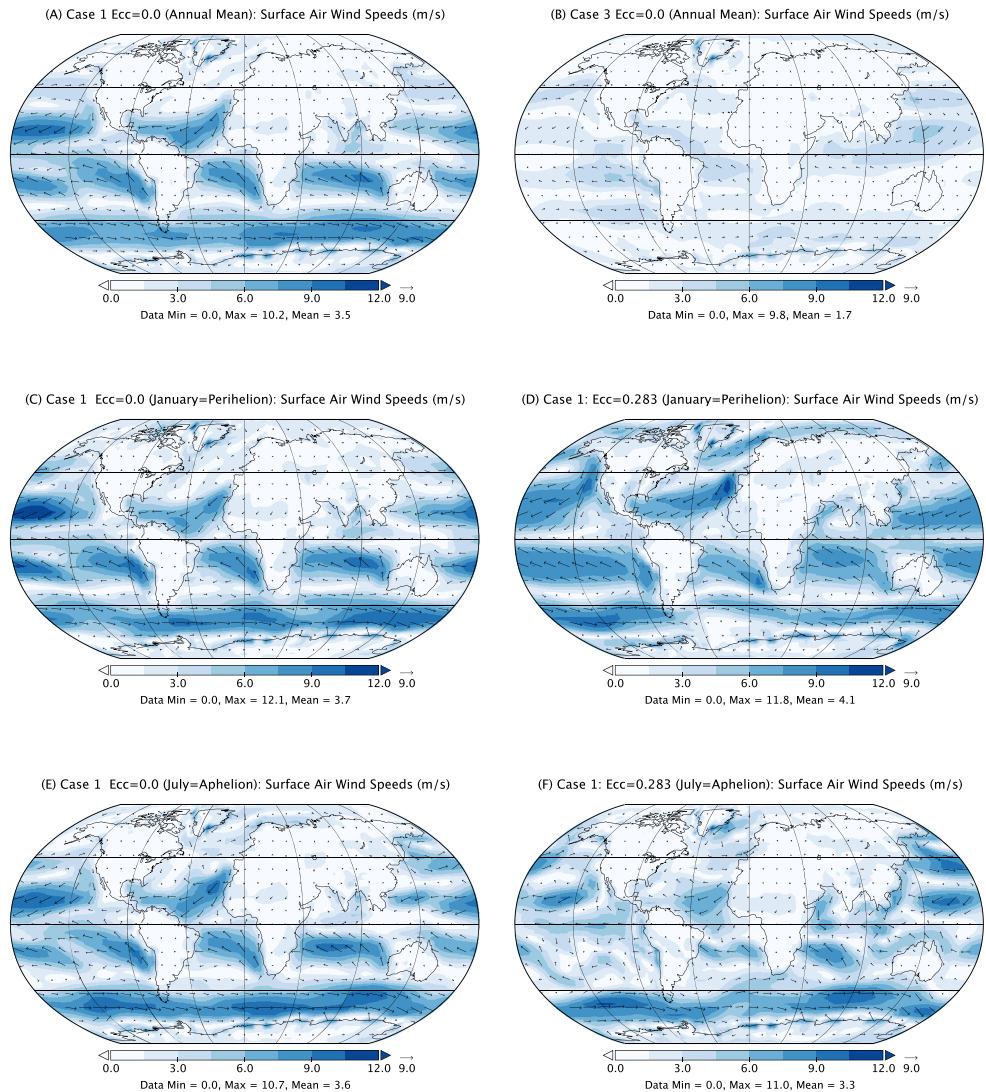

**Figure A6.** Mean wind speeds in meters per second. (A) is the annual mean for Case 1 (zero obliquity) when its eccentricity is at zero. (B) is Case 3 (45° obliquity) when its eccentricity is at zero. (C) is Case 1 in the month of January when the planet is at perihelion in the Southern Hemisphere summer. In the Southern Hemisphere, there tends to be more ocean than land where wind differences are more easily seen. (D) is meant to contrast with that of (C) where Case 1 now has an eccentricity of 0.283 demonstrating that the mean wind speeds are not markedly different. The same applies in plots (D) and (E) at aphelion/July.

### ORCID iDs

M. J. Way ⬛ https://orcid.org/0000-0003-3728-0475
Nikolaos Georgakarakos ⬛ https://orcid.org/0000-0002-7071-5437
Thomas L. Clune ⬛ https://orcid.org/0000-0003-3320-0204